\documentclass[10pt,a4paper]{article}
\usepackage[utf8]{inputenc}
\usepackage[english]{babel}
\usepackage{amsmath}
\usepackage{dsfont}
\usepackage{physics}
\usepackage{amsfonts}
\usepackage{amssymb}
\usepackage{mathtools}
\usepackage{graphicx}
\usepackage{slashed}
\usepackage{circuitikz}
\usepackage{comment}
\usepackage[left=2cm,right=2cm,top=2cm,bottom=2cm]{geometry}
\usepackage{caption}
\usepackage{subcaption}
\usepackage{makecell}
\usepackage{bm}
\usepackage{tikz}
\usepackage{axodraw2}
\usepackage{hyperref}

\usepackage[compat=1.1.0]{tikz-feynman}
\usepackage{cite}

\numberwithin{equation}{section}

\definecolor{DGreen}{rgb}{0.0, 0.42, 0.24}
\definecolor{OGreen}{rgb}{0.0, 0.52, 0.20}

\newcommand{\bit}{\begin{itemize}}
\newcommand{\eit}{\end{itemize}}
\newcommand{\ben}{\begin{enumerate}}
\newcommand{\een}{\end{enumerate}}
\newcommand{\beq}{\begin{equation}}
\newcommand{\eeq}{\end{equation}}
\newcommand{\bea}{\begin{eqnarray}}
\newcommand{\eea}{\end{eqnarray}}
\newcommand{\eean}{\nonumber\end{eqnarray}}

\newcommand{ \gsim}{\mathrel{\vcenter
     {\hbox{$>$}\nointerlineskip\hbox{$\sim$}}}}
\newcommand{\gappeq}{\mathrel{\rlap {\raise.5ex\hbox{$>$}}
{\lower.5ex\hbox{$\sim$}}}}
\newcommand{\lappeq}{\mathrel{\rlap{\raise.5ex\hbox{$<$}}
{\lower.5ex\hbox{$\sim$}}}}

\newcommand{\dslash}{ \, \partial   \! \! \! / ~ }
\newcommand{\Dslash}{ \, D  \! \! \! \! / ~ }

\newcommand{\Aslash}{ \, A   \! \! \! / ~ }
\newcommand{\Vslash}{ \, V  \!\! \! \! \! / ~ }

\newcommand{\hateps}{\hat{\epsilon}}
\newcommand{\MSbar}{\overline{MS}}
\newcommand{\Lmod}{ \Lambda_{\rm mod}}

\def\a{\alpha}
\def\b{\beta}
\def\g{\gamma}

\def\m{\mu}
\def\n{\nu}
\def\r{\rho}
\def\s{\sigma}

\renewcommand*{\thefootnote}{\fnsymbol{footnote}}

\begin{document}

\begin{center}
	{\Large {\bf  Renormalisation and invariants for two U(1)s }}
	\vskip 20pt
	{\large 
		Sacha Davidson$^1$\footnote{E-mail address:
			s.davidson@ip2i.in2p3.fr}, and Martin Gorbahn$^2$\footnote{E-mail address: martin.gorbahn@liverpool.ac.uk} }
	
	\vskip 10pt  
	
	{\it $^1$Université Claude Bernard Lyon 1, CNRS/IN2P3, IP2I Lyon, UMR 5822, Villeurbanne, F-69100, France
	}\\
    {\it $^2$Department of Mathematical Sciences, University of Liverpool, Liverpool L69 3BX, United Kingdom
    }\\
    
\end{center}
\begin{abstract}
\noindent

We revisit the renormalisation  of models with   two U(1) gauge symmetries, in a formulation  with non-canonical gauge kinetic terms which is covariant under field reparametrisations  among the two gauge bosons. 
This  approach is convenient  to study the appearance of kinetic mixing  in scale evolution, because  a coupling matrix is attributed to the gauge kinetic terms. 
We obtain simple $\overline{MS}$ renormalisation group equations up to two-loop,  which can be solved  to give effective millicharges  at low energy  which depend on the running couplings and  heavy mass scales of the model.
This formulation allows to construct ``invariants'' out of running Lagrangian parameters,  which are  invariant under generic gauge field reparametrisations, including rescalings, and  which { can be}  related  directly to observables such as  the millicharge.

\end{abstract}

\renewcommand*{\thefootnote}{\arabic{footnote}}
\setcounter{footnote}{0}

\section{Introduction}
\label{sec:intro}

  In the presence of more than one U(1) gauge symmetry, the gauge bosons
can mix in propagation. This is commonly parametrised as
 a  non-canonical kinetic term 
\beq
 {\cal L} \supset -\frac{\varepsilon}{2} F_{\mu \nu} V^{\mu \nu}  
\label{eqn1}
\eeq
where $F^{\mu\nu}$ and  $V^{\mu\nu}$ are the field strength tensors of two U(1)s, and  $\varepsilon$ can be induced  in renormalisation group running by vacuum polarisation diagrams involving heavy particles charged under both U(1)s (see Figure \ref{fig:loops}). 
As discussed by Okun\cite{Okun:1982xi} and  Holdom\cite{Holdom} long ago, this    ``kinetic mixing''  can be reparametrised into  ``millicharges'' for  the  light  particles in the model.
This setup is a popular example of light New Physics scenarios,
 which is constrained by many experimental searches  \cite{Golowich:1986tj,Prinz:1998ua,BABAR:2021cdg,Belle-II:2022jyy,BESIII:2022oww,NA64:2023wbi,NA64:2023ehh,FASER:2023tle,NA62:2023nhs}, and  whose astrophysical, experimental and cosmological  consequences have  been  reviewed in several studies
\cite{JR,SHIP,MATHLUSA,Beacham:2019nyx,Fabbrichesi:2020wbt}.
Extra light U(1)s can  impersonate diverse scenarios:
they  could be  the harbinger of a mirror universe \cite{Foot:2004pa, Berezhiani:2000gw},
 constitute cold dark matter if massive  and produced like axions during inflation \cite{Graham:2015rva} 
or via the misalignement mechanism \cite{Nelson:2011sf,Arias:2012az},
 and could modify thunderstorms \cite{Dmitrieva:2026ges}.

  This project has two aims; first,   to  renormalise a toy model with two U(1) gauge symmetries, in order to understand how kinetic mixing fits into renormalisation, and   address some naive confusions  about the identity of kinetic mixing (is it a  finite model prediction?
  A running parameter?)   and the behaviour of the couplings (do they mix as they run?  Do the charges run?).
  The second aim 
was to 
  construct  invariants (à la Jarlskog)  that correspond to physical observables, irrespective of the parametrisation chosen in  the Lagrangian.

  The renormalisation\footnote{  The impact of finite kinetic mixing has also been discussed for two
  \cite{Babu:1997st}
  and three U(1)s \cite{Heeck:2011md} (see also \cite{Rizzo:2012rf},  and references therein, for  an  application to   supersymmetric Grand Unified Theories). }   of  models with two U(1)s has been studied    carefully \cite{delAguila:1988jz,delAguila:1995rb,FFFT} for  a Lagrangian with canonical kinetic terms.
Reference \cite{delAguila:1988jz} obtained Renormalisation Group Equations (RGEs) at one and two-loop for  a triangular matrix of U(1) gauge couplings.
The subtraction of divergences was discussed in   \cite{delAguila:1995rb,FFFT}, where it was observed that there are  more divergences than U(1)s. It was nonetheless shown in  \cite{FFFT}, that if one refrains from introducing  off-diagonal kinetic counterterms because the tree-level kinetic terms are diagonal, the  theory is still renormalisable. 
  Luo and Xiao\cite{Luo:2002iq,Luo:2002ti}  included   non-canonical kinetic terms in multi-U(1) models, and generalised the  two-loop RGEs  of  Machacek and Vaughn \cite{Machacek:1983tz,Machacek:1983fi,Machacek:1984zw}
 with these terms. This generalisation is discussed on more detail in  \cite{Fonseca:2013bua}, where canonical kinetic terms are prefered, so the kinetic mixing is included in the RGEs via a  matrix of  couplings as in  \cite{delAguila:1988jz,delAguila:1995rb}.  
 The results   of  Luo and Xiao have been included  in  public codes such as  PyR@TE \cite{Lyonnet:2016xiz,Lyonnet:2016qyu} and RGBeta\cite{Thomsen:2021ncy}.

 We  focus on obtaining and solving RGEs because we are interested in renormalisation as scale evolution. 
 {   Since field reparametrisations are widely used  to modify the form of the Lagrangian, we aim for a formulation which  we call ``covariant'', meaning  the expressions remain valid under   rotations and rescalings  among the two U(1) gauge fields. }
 Such  a covariant  formulation  allows to   obtain  observables expressed in terms of ``invariants''. 
  The Lagrangian will be constructed with non-canonical kinetic terms, that is, with   a coupling matrix  containing  the (gauge coupling)$^{-2}$  multiplying  the  kinetic terms. 
This is  appropriate, 
because 
the physics of interest  is the appearance of kinetic mixing  during  scale evolution, and because this formulation is intrinsically covariant under field reparametrisations. 
Attributing a gauge coupling matrix to the kinetic terms  
 gives a clean separation between the running  gauge couplings, which appear in the  gauge kinetic terms, and the  conserved charges of matter particles, which appear in the covariant derivative.
  In this respect, our Lagrangian differs from that of  of Luo and Xiao\cite{Luo:2002iq}, where the gauge couplings stay in the covariant derivatives and
non-canonical kinetic terms  are added.

 We obtain an invariant  for kinetic mixing,  constructed from Lagrangian parameters, and invariant  under gauge field reparametrisations.
  It corresponds to the coupling combinations entering $S$-matrix elements,
  and is expressed in terms of running ($\MSbar$) parmeters evaluated at the experimental scale. 
  We give a covariant formulation for the  Renormalisation Group Equations (RGEs), which are linear at  one-loop where  the solution is  two independent  running couplings.
  The  third parameter  of the kinetic matrix, an angle, is scale-independent  at one loop, and runs at two-loop.
{ We also obtain (in Appendix \ref{app:invRGE}) the RGEs for invariants formulated in terms of invariants.}

Section \ref{sec:notn}  presents various forms of the  Lagrangian,  discusses the  field reparametrisations which make them equivalent, and introduces an ``invariant'' for kinetic mixing.
The  $\MSbar$ renormalisation, and the RGEs up to two-loop  are reviewed in Section \ref{sec:renorm}, in a formulation that is covariant under (scale-independent) field reparametrisations.  
The RGEs are solved  bottom up and top down in  Section \ref{sec:EFT},  which allows to check that the invariant of Section \ref{sec:notn} corresponds to the  effective millicharge.
We discuss and summarise in Sections \ref{sec:disc} and \ref{sec:sum}.

\section{ Notation and Field Reparametrisations}
\label{sec:notn}

The Lagrangian  of a  model  with two U(1) gauge symmetries
can be  formulated  in various ways. We will prefer Lagrangians with non-canonical kinetic terms, because  kinetic mixing is an interesting  aspect  of such models, and because renormalising the kinetic matrix is simpler than renormalising the couplings.
{   This sections starts by recalling   field reparametrisations   and how they motivate invariants in Section \ref{ssec:reparam},   discuss  Lagrangians   for two U(1)s in 
  Section \ref{ssec:L}, and  presents  the kinetic mixing invariant in Section
\ref{ssec:invarL}.}

\subsection{ Symmetries, reparametrisations and basis choices}
\label{ssec:reparam}

Symmetries and  field redefinitions
 can be defined in the context of the Path Integral. We restrict to the Lagrangian, on the   glib assumption that  modifications to the Path Integral measure (resulting from field redefinitions)   are cancelled out   of $S$-matrix elements by dividing by the  vacuum-to-vacuum transition amplitude. 

 A {symmetry} is a  transformation of the fields and/or coordinates,  under which the Lagrangian  remains {\it unchanged}---for instance, a global phase rotation of the scalar field in $\phi^4$ theory.  
 Global symmetries have associated conserved currents\footnote{Obtaining conserved currents is very similar to obtaining Lagrangian equations of motion(EoM), just that for the EoM, a total divergence is dropped  because the field variation vanishes at $\infty$.
 Whereas in the case of a symmetry, the Lagrangian is invariant under the field variation, which does { not}  vanish at $\infty$, so the  total divergence must vanish.}, which give quantities that are conserved by the { dynamics} following from the equations of motion.
 In perturbative  Quantum Field Theory,   the symmetries of  an interaction term ($eg$ phase rotations of charged particles) are  transfered to the Feynman rules, and states can be labelled by the eigenvalues of the generators of the symmetries ($eg$ the charge).

{ Reparametrisations  are redefinitions of the fields where  the Lagrangian parameters change  in compensation,  but  observables are invariant. In the Path Integral,   the fields are  integration variables, so can be redefined  provided that   the integrand  remains the same.
This is the Quantum Field Theory version of variable changes in usual integration, such as
$$
\int dA \exp\{-KA^2 +vA\} = \int \frac{d\tilde{A}}{\sqrt{K}} \exp\{ -\tilde{A}^2 +\frac{v}{\sqrt{K}}\tilde{A}\}~~~,~~~\tilde{A} =\sqrt{K} A
$$
So  in the words of Weinberg (Chapter 7.7 of Ref.~\cite{Weinberg:1995mt}), such field redefinitions are associated to  redundant  Lagrangian parameters. 
Indeed, reparametrisations are 
regularily used to  make the kinetic terms canonical in Lagrangians (see {\it eg} \cite{Babu:1997st,Davidson:2007si}
for finite reparametrisations in multi-U(1) models).}

 For some Lagrangians with canonical kinetic terms, there remain unitary transformations on the fields and parameters, under which the Lagrangian and the observables are invariant ---for instance, rotations in quark flavour space in the Standard Model.
 In these cases, one must choose a basis to calculate, but  it can be convenient to construct ``basis-independent'' ``invariants'' out of the Lagrangian parameters.
{ This has been performed for various models\cite{Davidson:2005cw,Davidson:1997mc,Davidson:1996cc};} a  well-known  example is the Jarlskog invariant \cite{Jarlskog:1985ht,Bernabeu:1986fc} (see \cite{Branco:1999fs} for a  detailed construction of   such CP-sensitive   invariants).

{ The ``invariants'' obtained here  differ from previous  constructions, in  that are invariant under  arbitrary gauge  field redefinitions, rather than just under the  residual rotations  consistent with canonical kinetic terms.}
 This is a simple generalisation, because invariants can be  associated to diagrams, and propagators have coupling constants when the kinetic terms are non-canonical.  It is warranted for this model, because kinetic mixing is a non-canonical kinetic term, and the various formulations of the model differ by rescalings as well as rotations among the gauge fields.

\subsection{Lagrangians with two U(1) gauge symmetries}
\label{ssec:L}

This section   discusses  different versions of the Lagrangian for our model. They differ by field reparametrisations, so should   give equivalent $S$-matrix elements. Our aim is to study the renormalisation of the model,  for which we need the bare  Lagrangian, but  it is convenient to  first choose the parametrisation of the fields, because it affects the $Z$-factors.

The  Lagrangians  describe a toy model  containing Dirac fermions  with  two  U(1) gauge symmetries.
However,  the U(1) of the Standard Model (SM) is chiral hypercharge,  so  a more realistic model  (see $eg$ \cite{Feldman:2007wj}) would  be subject to  anomaly cancellation conditions, and would have different numerical factors.
We envisage that these differences have little impact on our aims.
 
To fix notation we will consider the gauge bosons of two U(1)s assembled into a vector
\begin{displaymath}
  \vec{A}^\mu = ({A}^\mu, V^\mu)~~~.
\end{displaymath}
Both  are massless  for most of  our discussion (masses are discussed, for instance in \cite{Pospelov:2008zw}).
A Lagrangian can be written, for an arbitrary parametrisation   of  the gauge bosons  (who therefore wear primes), as
\beq
    {\cal L}_{nc} = -\frac{1}{4}({F}^{\mu\nu '}  V^{ \mu \nu '})
  \left[\begin{array}{cc}
K_{11}     & K_{12} \\
        K_{12} &K_{22}\end{array}
      \right]
  \left(\begin{array}{c}
{F}^{'}_{\mu\nu }\\  V^{'}_{\mu \nu}
  \end{array}
  \right)
  + \sum_f \overline{f} (i\Dslash^{'}-m_f) f + \mathcal{L}_{gf}
  + \mathcal{L}_{neglect}
\label{Lnc}
\eeq
where the gauge bosons interact  with  fermions $f$ via the covariant derivative
\beq
D^{\mu '} = \partial^\mu + i \vec{v}'\cdot\vec{A}^{\mu '}~~,
\label{Dcovar}
\eeq
with
$ \vec{v}'\cdot\vec{A}' = 
Q'_{f} A^{\mu '}  + Y'_{f } V^{\mu '}$, { so there  are no gauge couplings in the covariant derivative\footnote{ This  differs from the Lagrangian of \cite{Luo:2002iq}, where there are gauge couplings in the covariant derivatives  and non-canonical kinetic terms.}.}
The gauge kinetic terms can be written more compactly as $-\frac{1}{4}\vec{F}^{\mu\nu '}  [K] \vec{F}^{'}_{\mu \nu}$, where hermiticity and positivity imply that $[K]$ is a real symmetric positive definite matrix.

{  The  fermions $\{f\}$ will later be divided into   $\eta_j$s  of
mass $m_j$ which are charged under both U(1)s,  SM $\psi$s which   interact principally with $A^\mu$, and  shadow  $\chi$s  which interact principally with the  shadow photon   $V^\m$. And the terms in ${\cal L}_{neglect}$ which we neglect are sketched in Appendix \ref{app:neglect}.}

For the gauge-fixing term we choose $\mathcal{L}_{gf} = -\tfrac{1}{2}\partial_\mu \vec{A}^{\mu } [\Xi] \partial_\nu\vec{A}^{ \nu}$, where $[\Xi]$ is real and symmetric. In the following we will choose $[\Xi]=[K]$ for the renormalised Lagrangian, which  { corresponds to} Feynman Gauge  for  multiple U(1)s  and  our choice of the kinetic term.{  This is peculiar, because  $[K]$ renormalises and  $[\Xi]$ does not; however, the same peculiarity arises in   
  Feynman gauge with canonical kinetic terms, where $\xi = 1$ is not preserved by the renormalisation group equations.}
Yet this is inconsequential for all on-shell matrix elements and for the running or $[K]$ and $m_f$, as those are all independent of $[\Xi]$.

{ This  gauge-fixing choice} is symmetric-in-U(1)-space, so suitable for massless bosons. This can be compared to Ref.~\cite{FFFT}, which gives
a detailed discussion of gauge-fixing two U(1)s (in the canonical Lagrangian formalism) where the second U(1) is massive, with  distinct gauge choices for the two  U(1)s. We envisage that a gauge boson mass  $m_V$ can be neglected in the scale evolution of the model at scales  above $m_V$, so our results should apply, but that at the scale $m_V$, the massless model should be matched onto a massive one, where  a different gauge-fixing scheme could be chosen.

We find the above normalisation of the gauge kinetic term advantageous when discussing the renormalisation of the theory:
The U(1) Ward identities guarantee that $\vec{v}_f'$, $\vec{A}$ and $[\Xi]$ do not renormalise,  so only $[K]$ and the fermion masses $m_f$ run.

{ 
The form of the  Lagrangian  can be changed by  gauge fields reparametrisations, without affecting observables.   
  Since renormalisation will be discussed in  Section \ref{sec:renorm}, it is useful to  
  distinguish  whether these  gauge field transformations
  depend, or not, on the  renormalisation scale $\mu$.}

  Consider first the  case of  $\mu$-independent  changes  to  ${\cal L}_{nc}$, which will later serve  to  simplify the initial conditions  of the RGEs of this Lagrangian.
The kinetic matrix   $[K]$ can be diagonalised   with an orthogonal rotation $O_K$:
$$O_K [K] O_K^T=    [D_K]~~,$$
where the eigenvalues $1/e^2$ and $1/g^2$  of $[K]$ correspond to the gauge couplings  when the charges are defined as $(Q_f,\tilde{Y}_f) = (Q'_f,Y'_f)  O_K^T$. {This diagonalisation is motivated  because the external states in perturbative $S$-matrix calculations are the eigenstates of the  quadratic terms in the  Hamiltonian.} 
A further  reparametrisation of the gauge fields
allows to equate the eigenvalues of the kinetic terms (at tree level, or at some fixed scale in the renormalised theory):
we redefine $V^\mu \to \frac{e}{g} V^\mu$, so that
$\tilde{Y}_f \to Y_f = \frac{g}{e}\tilde{Y}_f$ and the diagonal
kinetic matrix becomes
\bea
    [D_K] = \left[\begin{array}{cc}
        \frac{1}{e^2}& 0 \\
        0 &\frac{1}{g^2}\end{array}
      \right]~~~ g^2 = e^2
\label{Ktree}
\eea
which  is then  invariant under O(2) rotations in the space of the two U(1)s. So imposing diagonal kinetic terms is not a basis choice. 
The charge vectors corresponding to this kinetic matrix are
\bea
\vec{v}_f=
\left(\begin{array}{c}
        Q_f \\
        Y_f\end{array}
        \right) = 
         \left[\begin{array}{cc}
         1& 0 \\
        0 &\frac{g}{e}\end{array}
           \right]
{\Big [} O_K {\Big ]}
         \left(\begin{array}{c}
        Q'_f \\
        Y'_f\end{array}
        \right)
        \label{vecv}
\eea

 Consider now renormalisation-scale dependant  gauge field redefinitions, which allow to change the form of the Lagrangian.  An interesting transformation is to diagonalise $[K]$ and   absorb the eigenvalues of $D_K$ into the gauge  fields:
  \bea
  \vec{A}_{can}^\mu = [D_K]O_K \vec{A}^{\mu '}~~~,
  \label{cannc}
  \eea
  which induces  canonical kinetic terms in the Lagrangian:
\bea
{\cal L}_{can} &=& -\frac{1}{4}\vec{F}_{can}^{\mu\nu }  \vec{F}_{can,\mu \nu} 
+ \sum_{f} \overline{f} (i\Dslash_{can}-m_{f}) f
\label{Lcan}%
\eea
where the  covariant derivatives now contain the couplings,
$
D_{can}^\mu = \partial^\mu + i\vec{u}_f \cdot\vec{A}_{can}^\mu$,
with  $\vec{u}_f = (Q_f e, \tilde{Y}_f g)$.

  Another common formulation of the Lagrangian with two U(1)s, is to
  add Eqn (\ref{eqn1}) to  a Lagrangian with canonical kinetic terms, where the
  matter particles $\{\psi\}$ interact with $A^\m$, and the $\{\chi\}$ interact with $V^\m$:
  \bea
  {\cal L}_{other}\!\! &\! =\!& -\frac{1}{4}{F}^{\mu\nu }{F}_{\mu \nu}
  -\frac{1}{4}{V}^{\mu\nu }{V}_{\mu \nu}
  -\frac{\varepsilon}{2}{V}^{\mu\nu }{F}_{\mu \nu}
  + \sum_{\psi} \overline{\psi} (i\dslash -eQ_\psi \Aslash  )\psi
  + \sum_{\chi} \overline{\chi} (i\dslash -eY_\chi \Vslash  )\chi ~~~~~~~~
  \label{Lother}
  \eea
  By making a field reparametrisation, such a Lagrangian can be    transformed  into  either of the   Lagrangians given in    Eqns (\ref{Lnc}),(\ref{Lcan}).
 The parameters of the various Lagrangians can be compared using the invariants  discussed in Section \ref{ssec:invarL}.

{
 Finally, we give our conventions for the relative normalisation of couplings and charges.  
  The   gauge  couplings parametrise the  relative importance of interactions vs propagation for the gauge fields,   
  whereas charges are required for  $\geq 2$ matter particles, to modulate the relative strength of their  gauge interactions.
  These two roles are  clearly separated   in the non-canonical  Lagrangian of Eqn (\ref{Lnc}), but the relative normalisation can be changed  ($Q\to xQ$, $e\to e/x$) by a field definition $A^\m \to A^\m/x$.
We  often  address this redundancy by choosing  $e = g$ and   $ Q_\tau = -1$   at some scale.
}

\subsection{A kinetic mixing invariant}
\label{ssec:invarL}

In this section, we  construct an invariant   for kinetic mixing, which corresponds to fermion scattering, and  is  unchanged under  generic  gauge field  redefinitions  including rescalings.  An invariant for a process with external gauge bosons is sketched in Appendix \ref{app:ttgg}. { Invariants for non-renormalisable interactions such as the Lamb shift could also be considered. }

\begin{figure}[htb]
   \begin{center}
\includegraphics[width=0.45\textwidth]{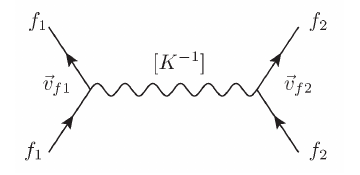}
\end{center}
  \caption{Tree-level scattering between fermions $f_1$ and $f_2$ in a 2 U(1)
    model parametrised by the  Lagrangian with non-canonical kinetic terms of Eqn (\ref{Lnc}). This diagram corresponds to the invariant of Eqn (\ref{ip}).
    \label{fig:milliinvar}
    }
\end{figure}

{ To construct  an invariant associated to Figure \ref{fig:milliinvar}, requires the  non-canonical  gauge boson propagator.  After subtracting total derivatives from the quadratic gauge boson terms in ${\cal L}_{nc}$ (Eqn \ref{Lnc}), and inverting,  the propagator is
\bea
-i [K]^{-1}\left\{ \frac{g^{\mu\nu}}{q^2} - \frac{q^\mu q^\nu}{q^4}\right\}
-i[\Xi]^{-1}  \frac{q^\mu q^\nu}{q^4} \to -i [K]^{-1} \frac{g^{\mu\nu}}{q^2} 
\label{prop}
\eea
where the Feynman gauge choice of $[\Xi] = [K]$ is implemented  after the arrow.
So the invariant corresponding to tree scattering amplitudes between particles $f_1$ and $f_2$ (see Figure \ref{fig:milliinvar}) is 
 \bea
I_{f_1 f_2} =  \vec{v}_{f1} \cdot [K^{-1}]\cdot \vec{v}_{f2} =
 \vec{u}_{f1} \cdot \vec{u}_{f2} ~~~.
 \label{ip}
 \eea
  where $\vec{v}_f$ are the charge vectors of Eqn (\ref{vecv}),   and  the (coupling constant)$^2$ { are contained in $[K]^{-1}$.} 
 The second  expression for the invariant is constructed from the Lagrangian with canonical kinetic terms; it is less useful than the first expression, because  it is not explicitly invariant under the field redefinitions that transform between the Lagrangians  of Eqn.s (\ref{Lnc}), (\ref{Lcan}) and (\ref{Lother}).  }

 Focusing on a single  shadow fermion $\chi$,
 the inner products  of Eqn (\ref{ip}) can be combined to obtain a  kinetic mixing or ``millicharge'' invariant,
 \beq
 I_\varepsilon \equiv
 \frac{\vec{v}_{\chi} \cdot[K^{-1}]\cdot \vec{v}_{\psi}}{\vec{v}_{\psi} \cdot[K^{-1}]\cdot \vec{v}_{\psi}}
\label{milliinvar}
\eeq
where $\psi$ is a SM fermion. 
$I_\varepsilon$ is invariant under  generic gauge field reparametrisations, so can be used to compare the parameters of Lagrangians  written  in different bases and with different field normalisations.
{ In a Lagrangian with
 canonical kinetic terms,} this invariant corresponds to the ``millicharge'' of  shadow particle $\chi$ normalised to the   SM charge
of $\psi$
 $$
 I_\varepsilon =
 \frac{Q_\chi}{Q_\psi}{\Big|}_{can} 
 $$
where  the SM fermions  $\psi$  interact exclusively with $A^\m$,  and the  shadow $\chi$s interact principally with the second U(1).
 In  the Lagrangian with non-canonical kinetic terms  of Eqn (\ref{Lother}), 
 $
 I_\varepsilon =
 -\varepsilon \frac{ Y_\chi g }{ Q_\psi e} {\Big|}_{other}
 $.

\section{Renormalisation}
\label{sec:renorm}

We are interested in renormalisation as scale evolution,  so  this section gives the RGEs  for our model up to two-loop.
We use dimensional regularisation, and  renormalise in $\MSbar$, so will later implement our RGEs in Effective Field Theory (EFT) in order to resum the correct logarithms \cite{Georgi:1993mps}.
The RGEs for the kinetic matrix $[K]$ are covariant in U(1) space, and linear at one-loop, so we will renormalise the non-canonical Lagrangian of Eqns (\ref{Lnc}) and (\ref{Ktree}). The results given here agree with \cite{Luo:2002iq} and \cite{Huber:2005ig}.

\subsection{The bare Lagrangian} 
\label{ssec:Lbare}

  In dimensional regularisation, the   bare Lagrangian in 4-$2\epsilon$ spacetime dimensions is 
  \bea
{\cal L}_0 &=& -\frac{1}{4}\vec{F}^{\mu\nu }  [K_0] \vec{F}_{\mu \nu}
  + \sum_f \overline{f}_0 (i\Dslash-m_{f,0}) f_0
   -\frac{1}{2 }\partial_\mu \vec{A}^{\mu } [\Xi_{ 0}] \partial_\nu\vec{A}^{ \nu}
  \label{Lbare0}
\eea
where the bare fields and parameters can depend on the regulator $\epsilon$, but should not depend on the scale $\mu$.  This  Lagrangian can be reexpressed in terms of  $Z$ factors and  running parameters   as
 \bea
{\cal L}_0 &=& -\frac{1}{4} \mu^{-2\hateps} \vec{F}^{\mu\nu }  [Z_K]^T[K][Z_K] \vec{F}_{\mu \nu}
    -\frac{1}{2 }\partial_\mu \vec{A}^{\mu } [{ \Xi}] \partial_\nu\vec{A}^{ \nu}
  + \sum_f Z_f \overline{f} (i\Dslash-Z_{m,f}m_{f}) f~~~,
 \label{Lbare}
 \eea
 where $D^{\mu} = \partial^\mu + i \vec{v}\cdot\vec{A}$, and  the running parameters    have the mass-dimension expected in 4-dimensions and can  be related to observations.
 The $Z$s contain the counterterms, which in $\MSbar$ can be expanded as a Laurent series in $1/\hateps$:
 \bea
Z = 1 + \frac{1}{\hateps}Z^{(1)} + \frac{1}{\hateps^2}Z^{(2)} +...
\label{Zdefn}
\eea
 where
 \beq
\frac{1}{\hat{\epsilon}}  \equiv \left(\frac{1}{{\epsilon}} -\g+\ln 4\pi\right)~~~.
\label{epshat}
\eeq
The $[Z_K]$ matrices will be  given in Eq.s (\ref{ZK1loop},\ref{invarRGEloops}).
Several comments are in order about the Lagrangian (\ref{Lbare}).

 Firstly,   the gauge fields  do not renormalise \cite{Tong}, and have mass-dimension one.
  This occurs because the vacuum polarisation diagrams  renormalise the kinetic matrix $[K]$  rather than the gauge fields. The usual Ward identities, such as $Z_1 = Z_2$ in QED,  are therefore more transparent.
{ In addition,  since  the Ward Identities imply that the gauge boson  counterterms are transverse, this implies that $[\Xi]$ is not renormalised, $Z_\Xi = 1$. As discussed after Eqn (\ref{Lnc}), we take $[K] = [\Xi]$ which corresponds to  Feynman gauge.}

 Secondly, the $Z_K$ factors are constructed  to be explicitly covariant in U(1) space, so $Z_K$ is a matrix:
  \bea
      [Z_K]^T[K][Z_K]&=&
      \left( \mathds{1} +\frac{1}{\hateps}\left[\begin{array}{cc}
          Z_{K,11}^{(1)} &         Z_{K,21}^{(1)} \\
          Z_{K,12}^{(1)}&        Z_{K,22}^{(1)}\end{array}
      \right] +..\right)
 \left[\begin{array}{cc}
     K_{11} & K_{12} \\
      K_{12}  &  K_{22}  
   \end{array}\right]
 \left( \mathds{1} +\frac{1}{\hateps}\left[\begin{array}{cc}
          Z_{K,11}^{(1)} &         Z_{K,12}^{(1)} \\
          Z_{K,21}^{(1)}&        Z_{K,22}^{(1)}\end{array}
   \right] +..\right)
 \nonumber\\
 &\approx & [K] + \frac{1}{\hateps} [Z_K^{(1)}]^T [K]
 +\frac{1}{\hateps} [K] [Z_K^{(1)}]
\label{ZK}
 \eea
where $[K]$ is symmetric and real, $[Z_K]$ is not required to be symmetric, and the +... are terms higher order in $1/\hateps$ which  are neglected.

{ 
This covariant formalism  for $[Z_K]$ differs from the common choice of writing a $Z$-factor for  each element of a matrix.
A first difference is that $[Z_K]$  has four matrix elements to implement the three counterterms that could be required for the symmetric $[K]$. We do not consider this to be a problem, because we later obtain a relation between elements of $[Z_K]$.
A second difference is that  our formalism allows a counterterm for $K_{12}$, even if $[K]$ is diagonal.  We consider this acceptable, because the rotation to diagonalise $[K]$ is $\mu$-dependent at $\geq 2$-loop, meaning that  the RGEs  regenerate  an initially vanishing $K_{12}$ (in  conventions where the charges  are $\mu$-independent).
 However, reference \cite{delAguila:1995rb} attached importance to the number of counterterms, and reference \cite{FFFT}  showed that,   in a model with a fermion, a massless and a massive gauge boson, observables could be renormalised without introducing a counterterm for $K_{12}$, but with mixing between the gauge couplings.

\begin{figure}[htb] 
  \begin{center}
  \includegraphics[width= 0.9\textwidth]{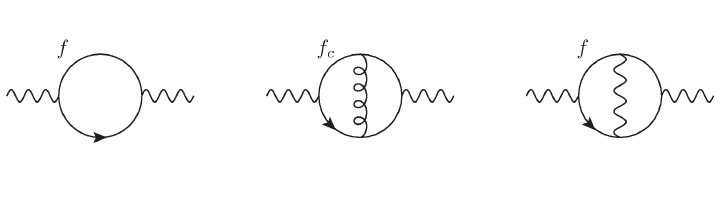}
\end{center}
\vspace{-1cm} 
  \caption{ The one loop  diagram and   representative two-loop  diagrams  contributing to vaccuum polarisation  in a 2 U(1)
    model containing various fermions $\{f\}$.  There are additional two-loop diagrams  with  the gauge boson vertices differently ordered on the fermion loop;
    only coloured fermions $f_c$ contribute in the QCD diagram.
    \label{fig:loops}.
    }
\end{figure}

The  $1/\hateps$ term of the $Z_K$ matrix will be required to obtain the RGEs.  It arises from  vacuum polarisation diagrams as  illustrated in Figure  \ref{fig:loops}.
For  canonical kinetic terms (and with the gauge choice $\xi = 1$), these diagrams are written as
\bea
\frac{-ig_{\m\r}}{q^2}
i[\Pi^{\r\s}] (q^2)
\frac{-ig_{\s\n}}{q^2}
\eean
where the  amputated bubble    $i[\Pi^{\r\s}]$    is a matrix in the space of the two U(1)s, commonly written as  $i ( q^2 g^{\mu \nu} -q^\m q^\n  ) [\Pi (q^2)] $, and presented in many textbooks. 
For  the  non-canonical Lagrangian,  the  coupling constants $e^2$ and $g^2$
{ are in the  amputated  external propagators (see Eq. \ref{prop})}, rather than in $[\Pi (q^2)]$. So for Dirac fermions at one loop, using Eqn (\ref{ZK})
\bea
Z_K^{(1)}{\Big |}_{1loop} = -\frac{1  } {24\pi^2} {\Big [}K^{-1} {\Big ]}
{\Big [}\sum_f \vec{v}_f \vec{v}_f^T{\Big ]}~~,~~
{\Big [}\sum_f \vec{v}_f \vec{v}_f^T{\Big ]} = \left[\begin{array}{cc}
     \sum_f Q_{f}^2 & \sum_f Q_{f} Y_f\\
      \sum_f Q_{f} Y_f &  \sum_f Y_{f}^2 
  \end{array}\right] 
\label{ZK1loop}
\eea
where there is an additional factor of 1/2  in the one loop coefficient for  chiral fermions.

 The two-loop contribution to the  matrix $Z_K^{(1)}$  is expected  to be numerically small compared to  one-loop (in our model where kinetic mixing arises at one loop; models where kinetic mixing first arises at higher order have been considered \cite{Gherghetta:2019coi}).
  However, it is formally interesting  because  it has  a different functional dependance on the couplings.
  Two-loop effects  have  been calculated  for QED \cite{Kallen:1955fb} (see also the summary in the  three-loop calculation \cite{Forner:2024ojj}), and we will include two-loop contributions  to the RGEs by generalising 
the QED $\b$-function  given in \cite{Huber:2005ig}. For this, we only need the ``covariants'' (matrices in the 2 U(1) space)  corresponding to the two-loop diagrams of Figure \ref{fig:loops}. Since the
amputated one-loop diagram gave the matrix  $  \sum_f \vec{v}_f\vec{v}^T_f$,  the two loop QCD and U(1) diagrams respectively give 
\bea
Z_K^{(1)}{\Big |}_{2loop} =
b_c g_s^2 {\Big [}K^{-1} {\Big ]}  {\Big[}\sum_c  \vec{v}_c \vec{v}_c^T{\Big]}  +
b_{U1} {\Big [}K^{-1} {\Big ]}  {\Big [}\sum_f   (\vec{v}_f \cdot [K^{-1}]\cdot \vec{v_f}) \vec{v}_f \vec{v}_f^T{\Big]}
\label{invarRGEloops}
\eea
where $b_c$ and $b_{U1}$ are numerical coefficents given at Eqn (\ref{RGEK}),  $g_s$ is the running coupling of QCD,  and $c$ indexes  coloured fermions.

\subsection{Renormalisation Group Equations}
\label{ssec:RGEsK}

This section derives  the RGEs for the kinetic  matrix $[K]$
(following the review \cite{Neubert:2019mrz}),  from which RGEs for the gauge couplings can be obtained.

The $\MSbar$ RGEs for $[K]$ can be obtained by imposing that the bare
$[K_0]$ of Eqn (\ref{Lbare0}) is  independent of the renormalisation scale $\m$:
 \bea
 0= \mu \frac{d}{d \mu} [K_0] &=& \mu \frac{d}{d \mu}
 \left(\mu^{-2\hateps }[Z_K]^T[K] [Z_K]\right)
 \eean
 where $Z_K$ is given in Eqns  (\ref{Zdefn},\ref{ZK}).  Below we use  $\mu \frac{d}{d\mu} = \frac{d}{d t}$.     Approximating $Z_K^{-1} \approx I$, this gives
  \bea %
  2\hateps [K] &=&
\left(\frac{d}{d t} [Z_K^T]\right)   [K] +\frac{d}{d t} [K]  
    + 
 [K] \left(\frac{d}{d t} [Z_K]\right)     ~~
 \eean
 where $\frac{d}{dt}[Z_K]$ is an expansion  of terms that diverge as $1/\hateps^n$  for $n\geq 1$, and recall that the only scale dependance in $[Z_K]$ is via the couplings. 
 Expanding the $\b_K$-matrix $\frac{d}{d t}   [K]$
 in positive powers of $\hat{\epsilon}$, as $ [\b^{(0)}   ] + \hateps
 [\b^{(1)}   ] +..$, it is straightforward to show  that  the $[\b^{(n)}]$ vanish for $n\geq 2$,  that $[\b^{(1)}   ]  = 2[K]$, and therefore that only
 $\frac{d}{dt}[Z^{(1)}_K]$ is relevant to determine the four-dimensional  $[\b_K^{(0)}]$:
   \bea
 [\b_K^{(0)}] =  -\left(\frac{d}{d t} [Z_K^{(1)}]^T\right) 
     [K]  
    -
 [K] \left(\frac{d}{d t} [Z_K^{(1)}]\right) ~~.
      \label{3.27}
      \eea
 Up to two-loops, from Eqns (\ref{ZK1loop},\ref{invarRGEloops}),
\bea
\frac{d}{d t} [Z^{(1)}_K]& = &  \left(\frac{d}{dt}[K^{-1}] \right) \left(- \frac{1  } {24\pi^2} {\Big [}\sum_f \vec{v}_f \vec{v}_f^T{\Big ]} 
+ b_c g_s^2  {\Big[}\sum_c  \vec{v}_c \vec{v}_c^T{\Big]}  +
b_{U1}   {\Big [}\sum_f   (\vec{v}_f \cdot [K^{-1}]\cdot \vec{v_f}) \vec{v}_f \vec{v}_f^T{\Big]} \right)
\nonumber\\&&
+
b_c \left(\frac{d}{dt}g_s^2\right) {\Big [}K^{-1} {\Big ]}  {\Big[}\sum_c  \vec{v}_c \vec{v}_c^T{\Big]}  +
b_{U1} {\Big [}K^{-1} {\Big ]}  {\Big [}\sum_f   (\vec{v}_f \cdot\left(\frac{d}{dt} [K^{-1}]\right)\cdot \vec{v_f}) \vec{v}_f \vec{v}_f^T{\Big]}
\eean
so using   $ \frac{d}{d t} [K^{-1}] = -[K^{-1}]
 \left(\frac{d}{d t} [K]\right)
[K^{-1}]$ and  the substitution 
 $\frac{d}{dt}[K]  \to [\b^{(1)}   ]  = 2[K]$ inside $\frac{d}{d t} [Z^{(1)}_K]$,
gives
\bea
[\b^{(0)}_K]& = &  - \left(\frac{1  } {6\pi^2} {\Big [}\sum_f \vec{v}_f \vec{v}_f^T{\Big ]} 
-8b_c g_s^2  {\Big[}\sum_c  \vec{v}_c \vec{v}_c^T{\Big]}  -
8b_{U1}   {\Big [}\sum_f   (\vec{v}_f \cdot [K^{-1}]\cdot \vec{v_f}) \vec{v}_f \vec{v}_f^T{\Big]} \right)
\eean
For a single U(1), the one-loop term reduces  to the QED  result,
so we extract $b_c$ and $b_{U1}$ comparing to the two-loop  QED RGEs of  \cite{Huber:2005ig}. This gives the $\b$-matrix for two U(1)s, up to two-loop,  which should be applicable in any basis:
\bea
\frac{d}{d t} [K] &=& 2\hateps [K] -  \frac{1}{6\pi^2}\left[\begin{array}{cc}     \sum_f Q_f^2 & \sum_f Q_f Y_f\\      \sum_f Q_f Y_f&\sum_f Y_f^2    \end{array} \right]  \nonumber\\
&& -
  \frac{\a_s(t) }{6\pi^3}\left[\begin{array}{cc}     \sum_c Q_c^2 & \sum_c Q_c Y_c\\      \sum_c Q_c Y_c&\sum_c Y_c^2    \end{array} \right]
-  \sum_f  \frac{\vec{v_f}\cdot [K^{-1}] \cdot\vec{v}_f}{32\pi^4}\left[\begin{array}{cc}  Q_f^2 &  Q_f Y_f\\    Q_f Y_f& Y_f^2    \end{array} \right]~~~~~~~~~~
\label{RGEK}
\eea
where the last two terms arise at two-loop,
$f$ indexes all  the fermions with masses below the scale of interest,
$\a_s$ is  the strong coupling,   and $c$ are coloured particles.

The one-loop   contribution to  the $\b$-matrix, which is the second term  on the right-hand-side of Eqn~(\ref{RGEK}), can be diagonalised by a scale-independent rotation,
implying that at one loop,  there is always a  basis choice  with two running couplings and constant charges. 
At two loop, so generically to all orders, the rotation to diagonalise the $\b$-matrix will depend on the renormalisation scale $\mu$ \footnote{ This agrees with   \cite{delAguila:1995rb}, where the coupling matrix was renormalised in the presence of canonical kinetic terms, and   three running parameters were expected. However \cite{delAguila:1995rb}  expected three running parameters already at one loop.}.
This $\mu$-dependance will be discussed in the next section; since  these RGEs are in a massless renormalisation scheme,  the impact of mass scales must be included by hand, and will be more important  that the running of  the diagonalisation angle.

\section{Solving the RGEs}
\label{sec:EFT}

In this section, we study the scale evolution of  the millicharge invariant,
by solving the RGEs for the kinetic matrix  after making a convenient choice of fields. { In Appendix
\ref{app:invRGE}, we obtain  and solve the one-loop RGEs for invariants 
in terms of invariants. }

It is notorious that dimensional regularisation, combined  with massless renormalisation schemes like $\MSbar$,  gives RGEs that  resum the  ``wrong'' logarithms.
However, implementing these RGEs in Effective Field Theory allows to recover  a more Wilsonian   scaling of parameters\cite{Georgi:1993mps}, relevant to observables.
So we transform our original model into a tower of sub-models, each defined in a scale slice between two mass threshholds of the  original model, to contain the particles which are light compared to the scale of the slice.  Then 
 we sequentially solve  the RGE of Eqn (\ref{RGEK}) passing through this  tower of  sub-models. 
Section \ref{ssec:bottomup}  runs the   RGEs bottom-up in a sub-model,  in order to show that the eigenbasis of the running couplings can in principle be observed.
Section \ref{ssec:milli} solves the one-loop RGEs top-down, from $\Lmod \to m_b$,   to show that   the invariant $\vec{v}_\psi[K^{-1}]\vec{v}_\chi$  is proportional to the  millicharge of a shadow fermion $\chi$.
Section \ref{ssec:2loop} includes two-loop QCD  contributions to the $\b$-matrix of the U(1) couplings, in order to explore the consequences of the  the $\b$-matrix  being scale dependent.

\subsection{Observing the  basis where the couplings  are  diagonal}
\label{ssec:bottomup}

We start at a  low  energy $\gsim m_b$,  in an EFT 
containing  the photon, the  paraphoton, the gluon, the SM fermions (except the $t$) and some  shadow fermions
$\{\chi\}$ lighter than the $b$.
This EFT  is defined between $m_b$ and $m_W$; at $m_W$ it matches to an EFT with more  SM particles, and at $m_b$,  it matches to an EFT without the $b$.
The aim is to explore what  can  be learned about this model from  observations
of  $\tau$s and one of the $\chi$s.
In particular,  is it in principle possible to observe  the angle $\theta_K$  diagonalising the RGEs, and  to observe   whether this angle is   scale-dependent?
 Associated to this question of scale-dependance of $\theta_K$, is  a notion  distinguishing  model parameters  that run,  from  parameters that  are $\m$-independent. We come back to this question in Section \ref{ssec:paramvspred}.

 At the scale $m_b$,  consider  scatterings among  $\tau$s and a particular  $\chi$.
  In a  { non-canonical}  gauge  field parametrisation such that the  kinetic terms are diagonal and 
  the photon is 
the gauge boson coupling to the $\tau$, one has
  \bea
 \vec{v}_\tau= \left(\begin{array}{c}
   -1\\
          0
    \end{array}\right)
~~~,~~~
 \vec{v}_\chi= \left(\begin{array}{c}
   Q_\chi\\
          Y_\chi
    \end{array}\right)
~~~,~~~
 D_K^{-1}(m_b) = \left[\begin{array}{cc}
   e^2&     0\\
         0&      g^2
     \end{array}\right]
  ~~{\rm with}~~~ e^2(m_b) = g^2(m_b)~. \label{bstart}
  \eea
{ Scattering the $\chi$s and $\tau$s with themselves and each other gives  three observables  evaluated at $m_b^2$, which } determine $e^2(m_b)$, $ Q_\chi$ and  $Y_\chi$:
\bea
\vec{v}_\tau [K^{-1}]\vec{v}_\tau (m_b) &=& e^2(m_b) \nonumber\\
\vec{v}_\tau [K^{-1}] \vec{v}_\chi(m_b)&=& -Q_\chi e^2(m_b) \nonumber\\ 
\vec{v}_\chi [K^{-1}]\vec{v}_\chi (m_b)&=& Q^2_\chi e^2(m_b)+ Y_\chi^2 g^2(m_b)
\eea

 The three observables  can be  translated  to the scale   $m_W$, by   running the RGEs  for
  $[K]$ (see Eqn \ref{RGEK}) upwards from $m_b \to m_W$.
  This  also  gives 
  \bea
 K^{-1}(m_W)& = &
 \frac{1}{\det\{K\}}
  \left[\begin{array}{cc}
    K_{22}(m_W)&     -K_{12}(m_W)\\
         - K_{12}(m_W)&       K_{11}(m_W)
    \end{array}\right]
  \nonumber\\
  &\approx&  
  \left[\begin{array}{cc}
   e^2(m_W)&     - e^2(m_W) g^2(m_W)K_{12}(m_W)\\
         -  e^2(m_W) g^2(m_W)K_{12}(m_W)&       g^2(m_W)
     \end{array}\right]
\label{K-2}
  \eea
where, in  the basis of eqn (\ref{bstart}), we allowed that the RGEs  could  induce off-diagonal kinetic terms for the gauge bosons, and, at one-loop
$$
K_{12} (m_W) = -\frac{1}{6\pi^2}
\sum_{ \chi} Q_\chi Y_\chi \ln \frac{m_W}{m_b}~~.
$$
The approximation for $K^{-1}$ relies on
$e^2(m_W) g^2(m_W)K_{12}(m_W) \ll 1$.
If the three observables are measured  at $m_W$ with arbitrary experimental precision,
they allow to determine 
$K^{-1}_{11}(m_W)\approx e^2(m_W)$,
$K^{-1}_{22}(m_W)\approx g^2(m_W)$, and $K^{-1}_{12}(m_W)\approx - e^2g^2K_{12}(m_W)$.
Measuring these three observables at some intermediate scale
  $q^2 \sim$ (40 GeV)$^2$, would  also allow to obtain 
$ e^2(q)$,  $g^2(q)$, and $K_{12}(q)$.

 It is in principle possible to  determine   whether  the eigenvalues of $[K]$  run with scale, and whether the angle to diagonalise it runs with scale,  by measuring   matrix elements of  $[K]$  with infinite precision at three  scales. To see this, start at some given scale, 
   for instance  $m_b$,   where $D_K$ can be chosen  $\propto I$. The  basis where the $\b$-fn is diagonal still exists, but  it cannot be identified from observations.
  So at $m_b$,  we take   $e^2 = g^2 =  |\vec{u}_\tau|^2$, and
measure $Q_\chi$ and $Y_\chi$ as above.
Then, as above,  measuring  $K^{-1}_{ij}(m_W)$ allows to solve for the eigenvalues and diagonalisation angle at $m_W$:
\bea
\lambda_{\pm K}= \frac{1}{2}{\Big\{} \Tr[K^{-1}] \pm \sqrt{\Tr[K^{-1}]  -4 \det [K^{-1}]}{\Big \}}
~~~,~~~
\tan 2\theta_K = \frac{2K^{-1}_{12}(m_W)}{K^{-1}_{22}(m_W) - K^{-1}_{11}(m_W)}
  \eea
{ Measuring  the three observables  at some other scale,  allows to verify whether  the angle  is scale-dependent, and thereby  whether there are two or three running parameters in $[K]$.}
This   applies  provided one does not cross particle threshholds.

\subsection{An expression for the millicharge in terms of running couplings}
\label{ssec:milli}

This section starts at a high scale $\Lmod$ where the model is defined,
and runs the kinetic matrix $[K]$  at one loop down to the low scale $m_b$.
The RGEs  are solved by running  through  a series of  sub-models,
which contain only the particles  which are lighter than the scales where the submodel is defined. 
This allows to evaluate, at $m_b$,   the millicharge  invariant of Eq. (\ref{milliinvar}).  

The particle content  of  the  model at $\Lmod$ is motivated by Holdom's study\cite{Holdom}: there are two massless gauge bosons, and  three groups of fermions (previously all called $f$): the  heavy $\{\eta_j\}$,  with $m_1>...>m_N> m_t$, are charged under both U(1)s,  $\{\psi\}$ are the Standard Model  fermions interacting principally with one of the gauge bosons, and the $\{\chi\}$ are some number of ``shadow'' fermions with masses $< m_b$ who interact principally with the orthogonal vector boson.
All the fermions are taken Dirac, although  $\{\psi\}$ and $\{\chi\}$ could be chiral. 

In  massless renormalisation schemes,  one  must implement by hand the mass threshholds  when solving the RGEs, in order that the  running  of the couplings resums the  leading log corrections to $S$-matrix elements.
This is a familiar recipe in EFT:  when running  down in scale,   every time  a mass threshhold is crossed,
 the massive particle is removed, and  the matrix elements of the models above and below the threshhold  are equated.  
We follow this recipe ``at one loop'', meaning that  the parameters run according to the one-loop RGEs (see Eqn \ref{RGEK}), the { tree}  matrix elements of $[K]$ are continuous  when crossing a threshhold $\mu = m$, and particles of mass $m$  are removed from the $\b$-function  at $\mu < m$.

At the high scale $\Lmod$, we suppose that    the $\psi$s and $\chi$s   do not interact at tree level: 
\bea
\vec{v}_\chi \cdot [K^{-1}] \cdot \vec{v}_\psi {\Big |}_{\Lmod} = 0~~~.
\eean
In order to solve RGEs, we need initial conditions for $[K]$ at $\Lmod$.
A convenient choice is to take $[K] \propto I$, as described in Eqn (\ref{Ktree}).
With this form of $[K]$,  which is then O(2) invariant, the $A^\m$ can be chosen in interact with the $\psi$s, and $V^\m$ interacts with the $\chi$s.  This basis choice, which  implies
\bea
\vec{v}_\psi = 
 \left(\begin{array}{c}
 Q_\psi  \\
 0
 \end{array}\right)~~~,~~~
 \vec{v}_\chi = 
 \left(\begin{array}{c}
 0  \\
 Y_\chi
 \end{array}\right) ~~~,
 \label{modelbasis}
\eea
is convenient because only the $\eta$s contribute to the off-diagonals of $[K]$.

Running $[K]$  down to the scale $m_1$  with the one-loop RGEs gives
  \bea
 \left[\begin{array}{cc}
   K_{11}&     K_{12}\\
         K_{12}&    K_{22}
     \end{array}\right](m_1)
 &=&  \left[\begin{array}{cc}
   \frac{1}{e^2}&     0\\
         0&        \frac{1}{g^2}
    \end{array}\right](\Lmod) 
   +  \frac{1}{6\pi^2}
    \left[\begin{array}{cc}
   \sum_f Q_f^2&     \sum_\eta Q_\eta Y_\eta\\
         \sum_\eta Q_\eta Y_\eta&       \sum_f Y_f^2
     \end{array}\right]\ln \frac{\Lmod}{m_1}
\eean
where  $f$ runs over all the fermions in the model,  $\eta$ runs over all the $N$
$\eta$s, and $g(\Lmod) = e(\Lmod)$.

At $m_1$, the full model is matched onto the submodel  constructed by removing $\eta_1$ from  the full model: the $K_{ij}$ are  taken continuous across the threshhold, and $\eta_1$ is removed from  the $\b$-function at lower scales.  Running down to $m_2$, this gives
\bea
 \left[\begin{array}{cc}
   K_{11}&     K_{12}\\
         K_{12}&    K_{22}
     \end{array}\right](m_2)
  &=&   \left[\begin{array}{cc}
   K_{11}&     K_{12}\\
         K_{12}&    K_{22}
   \end{array}\right](m_1)
    +  \frac{1}{6\pi^2}
    \left[\begin{array}{cc}
   \sum_f' Q_f^2&      \sum_{\eta = 2}^{N} Y_\eta Q_\eta\\
          \sum_{\eta = 2}^{N} Y_\eta Q_\eta &       \sum_f' Y_f^2
     \end{array}\right]\ln \frac{m_1}{m_2}
\eean
where the primed sums over  run over all the fermions $f$ except $\eta_1$.

Repeating this matching process across  all the threshholds down  to  the mass $m_b$, gives
  \bea
 \left[\begin{array}{cc}
   K_{11}&     K_{12}\\
         K_{12}&    K_{22}
     \end{array}\right](m_b)
&=&
   \left[\begin{array}{cc}
   \frac{1}{e^2}&     0\\
         0&        \frac{1}{g^2}
    \end{array}\right](\Lmod) 
   +  \frac{1}{6\pi^2}
    \left[\begin{array}{cc}
   \sum_f Q_f^2 \ln \frac{\Lmod}{\tilde{m}_f} &      \sum_\eta Y_\eta Q_\eta \ln \frac{\Lmod}{m_\eta}\\
         \sum_\eta Y_\eta Q_\eta \ln \frac{\Lmod}{m_\eta}&       \sum_f Y_f^2 \ln \frac{\Lmod}{\tilde{m}_f}
     \end{array}\right]
\label{K2mn}
    \eea
where in the logarithms, $\tilde{m}_f$ is $m_n$ for $f = \eta_n$,  $m_t$ for $f =t$,   and $m_b$ for  $f$ any SM or shadow particle of $m<m_b$.

The overlap  $\vec{v}_\chi\cdot K^{-1}\cdot \vec{v}_\psi$  parametrises the scattering between SM and shadow fermions. To evaluate this overlap  at the scale $m_b$, we  need to invert $[K](m_b)$, which can be performed analogously to Eqn
(\ref{K-2}).
\bea
\vec{v}_\chi\cdot K^{-1}\cdot \vec{v}_\psi (m_b) \approx  -
 e^2(m_b) g^2(m_b)Q_\psi Y_\chi K_{12} (m_b)
\label{milli3}
 \eea
 We calculated this invariant by remaining in  the  same basis  (given in Eqn \ref{modelbasis}) through the whole sequence of sub-models between the full model and low-energy. So the kinetic terms were not diagonal in the submodels, 
 but  this should not pose a problem because   we are calculating  an invariant.
 The millicharge invariant of Eqn (\ref{milliinvar}), which gives the millicharge of $\chi$  normalised to the charge of $\psi$, is therefore
  \bea
I_\varepsilon (m_b) \approx
 - \frac{g^2(m_b)}{6\pi^2} Y_\chi \sum_\eta Y_\eta Q_\eta \ln \frac{\Lmod}{m_\eta}
 \label{millichargef}
  \eea
  where 
  the charges on the right-hand side are in the high-scale model-basis of Eqn (\ref{modelbasis}).
As anticipated, the effect of RG running is to give a millicharge to $\chi$. 
In this one-loop expression,  the millicharge  is proportional to the running shadow coupling$^2$ (evaluated at the scale where the millicharge is probed).

\subsection{QCD at two-loop}
\label{ssec:2loop}

Two loop contributions to the running of $[K]$ are illustrated in Figure \ref{fig:loops}, and correspond to the  last two terms on the right side of Eqn (\ref{RGEK}).
They are  suppressed  relative to the one-loop term by  respectively $\sim \a_s/\pi$   and $\sim \a_e/\pi$, so are unlikely to be of numerical importance.
The QCD corrections are discussed here, because they make the $\b$-matrix scale-dependent, therefore the  rotation angle to diagonalise  $[K]$ could be $\mu$-dependent.
So at two-loop and beyond, $[K]$  will be  characterised by its two  eigenvalues (the  inverse coupling$^2$s) which already were $\mu$-dependant at one loop, and by a running angle.   
This section aims  to see the effects of   having a third running parameter  in $[K]$; however we do not perform a  full 2-loop analysis which   would require one loop matching calculations.

The two-loop QCD contribution can be  included by  implementing the one-loop solution to the RGE of $\a_s$:
\bea
\a_s(t) = \frac{\a_s(t_i)}{1+\a_s(t_i) \frac{\b_{s0}}{2\pi}\ln\frac{\mu}{\mu_i}}
\eea
with $\b_{s0}= (11-2N_{fc}/3)$ and $N_{fc}$ the number of colour triplet  Dirac fermions.  Integrating the right side of the RGE Eqn (\ref{RGEK}), where only the second and third terms are included, and using
$\int dt \a_s = \int d \a_s/\dot{\a}_s$, gives
\bea
 [K](\mu_f) &=&  [K](\mu_i)  -  \frac{1}{6\pi^2}\left[\begin{array}{cc}     \sum_f Q_f^2 & \sum_f Q_f Y_f\\      \sum_f Q_f Y_f&\sum_f Y_f^2    \end{array} \right]  \ln\frac{\mu_f}{\mu_i}
 +
  \frac{1 }{3\pi^2\b_{s 0}}\left[\begin{array}{cc}     \sum_c Q_c^2 & \sum_c Q_c Y_c\\      \sum_c Q_c Y_c&\sum_c Y_c^2    \end{array} \right]  \ln\frac{\a_{s}(\mu_f)}{\a_s(\mu_i)}~~~~~~~~~
\label{RGEKQCDsoln}
\eea
where $f$ runs over all the Dirac  fermions in the model, and $c$ runs over the coloured triplets.

The angle to diagonalise $ [K](\mu_f)$ is simple to obtain. For a model where only SM fermions $\psi$ are coloured,  in a basis where $\vec{v}_\psi = (Q_\psi,0)$  and  the  parametrisation  where
$ [K](\mu_i) \propto I$: 
\bea
\tan 2\theta &=& \frac{      \sum_f Q_f Y_f}{\sum_f(Y_f^2- Q_f^2) + \frac{2}{\b_{s0}} \sum_cQ_c^2 \frac{\ln [\a_{sf}/\a_{si}]}{\ln[\mu_f/\mu_i]}}
\label{thetaQCD}
\eea
where $\a_{sf} =\a_s(\mu_f)$. One sees  that  the angle is scale independant if QCD is neglected,  and that the angle vanishes if $\sum Q_f Y_f = 0$,  
which  corresponds to  no particles charged under both U(1)s, and
 SM particles interacting via QCD and QED (with a disjoint  shadow U(1)).
In the case where SM QCD corrections are included to a model with kinetic mixing, then
 $[\sum_f \vec{v}_f \vec{v}_f^T]$ and $[\sum_{c} \vec{v}_c \vec{v}_c^T]$ are not simultaneously diagonalisable, so  the eigenbasis of   $[K]$  rotates with scale, and $[K]$ is described by two $\mu$-dependent couplings  and a $\mu$-dependent angle.

The impact of a running angle in $[K]$  on $S$-matrix elements seems minimal.
Our aim in running RGEs is to resum leading log corrections into the couplings, such the calculation of  a tree-level matrix element gives an  leading-log improved result.
This resummation was obtained in the previous section  at one-loop,  and gave an effective millicharge depending on the running couplings at the experimental scale, and  logarithmically on the mass scales of the  full model.
So the effective charges at low energy are already  functions of $\mu$ at one-loop.
The form of the equation for the millicharge,  Eqn (\ref{millichargef}), is unchanged when  two-loop QCD is included for SM particles in the evolution from $\Lmod \to m_b$, but the expression for $e^2(m_b) $ is augmented by the two-loop QCD running from $\Lmod$.

\section{Discussion}
\label{sec:disc}

  \subsection{    Parameters,  predictions, and which is  kinetic mixing?}
  \label{ssec:paramvspred}

  One of the puzzles motivating this study was the identity of kinetic mixing. 
This Section argues that, as expected, it is a prediction of the model.

  The model is defined by its symmetries and particle content, and two types of   parameters. First are those  which should be input to the model at some scale, because  they  evolve with scale ($eg$ the couplings and masses in QED or QCD). 
  In  our model of 2 U(1)s parametrised with non-canonical kinetic terms, the coupling matrix has three parameters, and they all run.
  However, to set initial conditions for the RG evolution of the coupling matrix, only two eigenvalues are required.
  The scale evolution that generates the off-diagonal is a prediction of the model.

  Then there are parameters which are  scale-independent, so can be input to the model without a scale attached.  For instance, in  QED, the Ward-Takahashi identities  say that  electric charges are such parameters. 
  In a model with 2 U(1)s,  one can see that the charges are  scale-independent constants 
  by writing the Lagrangian with  non-canonical kinetic terms as in Eqn (\ref{Lbare}).
  In this formulation,  the gauge fields and charges  need no  renormalisation. This follows from the BRS identities, which  imply that the charges are scale-independent.

  At one-loop, the expression for the millicharge invariant is given in Eqn (\ref{millichargef}). 
  It  is a function of  the running coupling   $g^2(\mu)$, of charges, 
  and  of (running)  mass  parameters of the theory, such as the masses $m_i$ of the $\eta_i$,  or the scale $\Lmod$ where the model is defined.
  So modulo $\Lmod$ which we discuss next,   the invariant at low energy  is clearly a prediction  of the model, in the case where the  model's parameters were input at some higher scale.

  There has been some discussion in the literature (see $eg$ \cite{delAguila:1995rb}) about whether $\ln \Lmod$ should appear in  the formula for low energy kinetic mixing, because it suggests an undesirable dependance on  a cutoff.
  However, we are undisturbed by $\Lmod$,   because it either cancels out  in the sum over $\eta_j$  in the one-loop result for  the low-energy millicharge given in Eqn (\ref{millichargef}), or it is the physical scale of the theory
  where $\vec{v}_\psi [K]^{-1} \vec{v}_\chi (\mu) = 0$. (Recall that the $\eta$s
  are the  heavy fermions charged under both U(1)s, which generate the kinetic mixing.)

An input assumption of our model is that  the SM  and shadow fermions do not scatter  off each other  at the high scale $\Lmod$: $\vec{v}_\psi  [K]^{-1}\vec{v}_\chi (\Lmod)  = 0$. This assumption is
  equivalent to imposing that there is  a basis where $[K](\Lambda_{\rm mod}) \propto I$,  the SM particles have charges $Q_\psi$ but $Y_\psi= 0$, and the shadow particles have charges  $Y_\chi$ but $Q_\chi= 0$ (see Eqn \ref{modelbasis}).  If the $\b$-matrix is diagonal in this basis,  then $\vec{v}_\psi  [K]^{-1} \vec{v}_\chi= 0$  from $\Lmod\to m_1$, and  at one loop the off-diagonals of the $\b$-matrix vanish:
  $ 0 = \sum_{\eta_j} Q_j Y_j.$ So in this case, $\Lmod$ cancels out of the low energy millicharge (Eqn \ref{millichargef}).
  
  If the $\b$-matrix is {\it not}  diagonal in the $Y_\psi= Q_\chi=0$ basis at $\Lmod$, then the vanishing of  $\vec{v}_\psi [K]^{-1} \vec{v}_\chi$ at  $\Lmod$ is   an accidental cancellation that occurs only at $\Lmod$, which allows to calculate $\Lmod$ from the charges, masses and couplings of the model.

So in summary, to obtain predictions   in  our model  containing $N_f$ fermions,  one must input   the two couplings  and the masses of the $N_f$ fermions at some scale. In addition, the  2$(N_f-1)$ charges of the fermions must be specified.  This is all as expected.
One of the predictions of the model is     the millicharge invariant at low energy.

  \subsection{  If millicharges  run, what happened to current conservation and anomalies?}
  \label{ssec:currentcons}

  Current  conservation,   in classical electrodynamics, is a consequence of  the rephasing invariance of the charged particles, and applies in conjunction with  the  Equations of Motion.
  In quantum field theory,    where the Equations of Motion may not be satisfied,  { they can be exchanged  for}  Ward-Takehashi identities, which follow from a residual local symmetry that survives gauge fixing.
  These identities for our model can be obtained by making a BRS transformation  on  the non-canonical Lagrangian of Eqn (\ref{Lnc}) augmented by gauge-fixing terms as in  Eqn (\ref{Lbare}) and by ghosts.
  These identities confirm that the charges  do not run, as anticipated in Section \ref{ssec:Lbare}. Anomaly cancellation works as usual in this formalism, where the kinetic mixing  in the gauge boson propagators is  external to the triangle diagram.  { So one sees that, in the non-canonical formalism, anomalies cancel and   charge conservation identities  are satisfied as usual.}

{ The situation is less clear with the  canonical Lagrangian of Eqn (\ref{Lcan}),}
{ where  the millicharge of $\chi$ is proportional to
  the invariant $I_\varepsilon$  of  Eqn (\ref{millichargef}). 
 So already at one loop, there is a charge which  depends
   on scale via  the running couplings. However,}
the Path Integral is  invariant under field reparametrisations, and $S$-matrix elements are functions of reparmetrisation-invariant combinations of Lagrangian parameters. 
So the physical consequences of  Ward-Takehashi identities
and anomaly cancellation should remain in the canonical  parametrisation.
But it is reassuring 
to check that the effective charges in the canonical formalism are conserved at vertices, due to  the vanishing of the   sum of the charges of the fields participating in  each term of the Lagrangian.
Similarly,  the Hydrogen atom is neutral in the canonical formulation, because anomaly cancellation related the charges of the electrons and quarks in the non-canonical Lagrangian,  and  the proton and electron charge vectors   transform the same way in  going from the non-canonical  to canonical formulation.

  \subsection{ What about a covariant canonical formalism?} 
  \label{ssec:wrongwu?}

  The charge vectors $\vec{u}_f$  of the canonical Lagrangian (\ref{Lcan})  are physically attractive, because their inner products appear in $S$-matrix elements, and they are invariant under  rescaling of the charges into couplings.
  However, it is difficult to formulate and solve  covariant-in-U(1)-space  RGEs for  $\vec{u}$.
   We encountered two difficulties.
  
  The first is that the one-loop RGE for $e$ in QED is non-linear: $\mu \frac{d }{d\mu} e = \frac{1}{12\pi^2} e^3$.  So  the generalisation of this differential equation  to  the vector of couplings $\vec{u}$ is not obvious.
  This is to be compared with the one-loop RGE for the   kinetic term coefficient  $1/e^2$, which is linear and trivial to generalise to 2 U(1)s.

  The second issue, is that  rotations (in two-U(1)-space) of the  charge vector $\vec{u}$ are unrelated to the  rotation which diagonalises the kinetic terms.
  This can be seen from Eqn (\ref{cannc}), which implies that $\vec{u} = D_K^{-1}O_K \vec{v}$.
If the rotation to diagonalise  $[K]$ is scale-dependent, which  is not the case at one loop but appears true to all orders, then that rotation, given as $O_K$ in
  Eqn (\ref{cannc}),  is inconveniently  intricated inside $\vec{u}$. Furthermore,  $\vec{u}$  appears to contain two (running)  parameters, so its unclear what happened to the angle. 

 A more promising approach  that has been carefully explored \cite{delAguila:1988jz,delAguila:1995rb,Fonseca:2013bua},
 is to define a coupling matrix $[E]$, such that $\vec{u}_f =[E] \vec{v}_f$.  However, this requires defining $[E]$ as the square root of $[K^{-1}]$,
 and the commutator of  $[E]$ with its derivative seems required  to obtain its RGEs.  Therefore it appears more convenient to work with $[K]$.

  \subsection{Invariants}
\label{ssec:invar}

The invariants of Eq.s (\ref{ip}) and (\ref{milliinvar})  have many desirable features:  they are  invariant under rotations and rescaling of the U(1) gauge fields, they are  simple,   can be contructed diagrammatically,  and  they  correspond to the  coupling combinations that appears in $S$-matrix elements.

This is a rare combination. Most such constructions, including the familiar Jarlskog invariant,  are only  invariant under field rotations in a Lagrangian with canonical kinetic terms.
In addition, invariants constructed from the Lagrangian  are often only remotely related to $S$-matrix elements.
For instance, in the SM quark sector,  the Jarlskog invariant is  a trace (so corresponds to a closed diagram),
and is    constructed from tree-level parameters.
Whereas $S$-matrix elements measuring  quark CP violation do not correspond to closed diagrams, and are  proportional to simpler CP invariants and running parameters\cite{Brod:2019rzc}.
It has been suggested  \cite{Ardu:2024bua} that such  $S$-matrix combinations could be useful in reconstructing New Physics models;
however,  in our two U(1)  model,   the  Lagrangian  and $S$-matrix  invariants  are  the same.

An additional remarkable feature of the invariants $I_{f_i f_j}$ of Eq. (\ref{ip}), is that their RGEs can be expressed in terms of invariants (see Appendix \ref{app:invRGE}). This opens tantalising perspectives:  $S$-matrix elements  are functions of  invariants; if in addition,  RGEs can be written in invariant fashion, then perhaps  there is no need for the  Lagrangian parameters  deemed redundant by Weinberg \cite{Weinberg:1995mt}.

\section{Summary}
\label{sec:sum}

The Path Integral and Greens functions are invariant under field redefinitions --- but renormalisation-scale-dependent redefinitions can modify  which parameters run.
In a model with kinetic mixing between two U(1) gauge  bosons, we obtain  and solve  the $\MSbar$ RGEs (Eq. \ref{RGEK}) in a parametrisation where the kinetic terms have a running coupling matrix $[K]$, but there  are no U(1)  gauge couplings in the covariant derivatives.
We rediscovered that in  this parametrisation ( $\sim -\frac{1}{4e^2} FF$ for a single U(1)),  the gauge fields are not renormalised, the Ward identities are transparent, and  the RGEs are linear at one loop so trivial to solve.
$[K]$ at one loop   contains two running couplings and can be diagonalised by a finite angle.
This parametrisation is particularily convenient for multi-U(1) models,  because kinetic mixing is generated  automatically in RG evolution, and because the formalism  is covariant under  the $\mu$-independent field reparametrisations which are commonly used in such models.
To have a measure of kinetic mixing that is independant of the field parametrisation, we constructed a generalised invariant for kinetic mixing {(Eqn \ref{millichargef})}.
Recall that usual invariants ($eg$ Jarlskog) are { not modified by} unitary transformations  which leave the kinetic terms invariant in the Lagrangian; we generalised this notion  to { obtain an invariant which is unchanged } under generic gauge field reparametrisations including rescalings.

\subsection*{Acknowledgements}
SD thanks Andreas Crivellin  for enthusiastic discussions leading to this project.
The work of MG is supported by the UK Science and Technology Facilities Council
grant ST/X000699/1.

{ This project was finished at the Munich Institute for Astro-, Particle and BioPhysics (MIAPbP) which is funded by the Deutsche Forschungsgemeinschaft (DFG, German Research Foundation) under Germany´s Excellence Strategy - EXC-2094 - 390783311.
}

\appendix

\section{Neglected  Lagrangian terms}
\label{app:neglect}

This Appendix discusses the  various terms  not considered in the Lagrangian of Eqn (\ref{Lnc}).

Non-abelian  gauge interactions were not written because they are independent, at the Lagrangian level, of  the U(1)s. We also  neglected terms such as
  \bea
 \!\!\!\!   {\cal L}_{neglect}\!\! &\!=\!&\!\! -\frac{\varepsilon_{\m\n\a\b}}{2} ({F}^{\mu\nu '}  V^{ \mu \nu '})
  \left[\begin{array}{cc}
T_{11}     & T_{12} \\
        T_{12} &T_{22}\end{array}
      \right]\!
  \left(\!\begin{array}{c}
{F}^{\a\b'}\\  V^{\a\b'}
  \end{array}\!\!\!
  \right)
  + \sum_j D^{\mu '} \phi_j^\dagger D^{'}_\mu\phi_j - V(\phi_j) + (\lambda^{jkn}\phi_j \overline{f}_k f_n + h.c.)~~~~~~~
\label{Lnot}
\eea
where the Yukawa interactions are gauge invariant if the sum of the participating  charge vectors  vanishes: $\vec{v}_\phi' + \vec{v}_{fk}'+ \vec{v}_{fn}'=0$.
(This condition remains satisfied under rotations and rescaling in the  two-U(1)-space\footnote{In a particular field parametrisation,  the sum of the charges only needs to be an integer in order for the Yukawa interaction to be invariant. But the integer should be zero  to allow field reparametrisations which rescale the couplings $e\to e/x$ and charges $Q\to xQ$.}.)
Multiple scalars,  non-abelian gauge sectors  and Yukawa interactions are included in \cite{Luo:2002iq}.
Scalar fields   are neglected here (despite that they could induce  gauge boson masses), because we  are interested in the running of parameters  at scales above
possible gauge boson masses. 
Finally, in QED, the  CP-odd  $\theta$-term can be written as a total divergence, and this remains true for $T_{12}$ terms in the presence of 2 U(1)s
\bea
\varepsilon^{\m\n\a\b} {F}'_{\mu\nu }  V'_{ \a\b } &=&4 \partial_\mu (
\varepsilon^{\m\n\a\b}A'_\n\partial_\a V'_\b)~~~.
\eean
So we also neglect these $\varepsilon\vec{F}'[T]\vec{F}'$ terms, because they should be irrelevant for $S$-matrix elements, although they could be interesting in certain topologies or for space-time dependent $T_{ij}$\cite{Tong}.

\section{Invariant form of the RGEs}
\label{app:invRGE}

In the main body of the text we solved the renormalisation group equations for the coefficient of the kinetic term $[K]$, which then determined the running of the invariants.
Given the simple solution for the RGEs of $[K]$ this is a numerically efficient and practical approach, even though $K$ { transforms under field redefinitions}.
Hence, it is interesting to note that the RGE and their solution can be derived solely in terms of the invariants.
Writing $t=\ln \mu$ and sandwiching the RGE for $K$ inverse with the charge vectors of fermion $f_1$ and $f_2$
\begin{equation}
  \label{eq:ddtkinv}
  \frac{d}{dt} v_{f_1}^T [K]^{-1} v_{f_2} =
  - \frac{1}{6 \pi^2} v_{f_1}^{T} [K]^{-1} \sum_f v_f v_f^{T} [K]^{-1} v_{f_2}
  - \frac{\alpha_s(t)}{6 \pi^3} v_{f_1}^{T} [K]^{-1} \sum_c v_c v_c^{T} [K]^{-1} v_{f_2}\,,
\end{equation}
we find a RGE equation that is expressed only in terms of the invariant matrix  defined in Eq.\eqref{ip} (Note that we sum over all fermions $f$, treating each coloured state distinct, and only over the coloured triplets $c$, when we include mixed QCD effects).
Using this equation, we find for the inverse of the invariant matrix 
\begin{equation}
  \label{eq:inverse-I-ODE}
  \frac{d}{dt} [I]^{-1} =   [I]^{-1} [\dot{I}] [I]^{-1} =
- \frac{1}{6 \pi^2} \mathds{1} - \frac{\alpha_s(t)}{6 \pi^3} \delta_c\,,
\end{equation}
where $\mathds{1}$ denotes the unit matrix in fermion space, while $\delta_c$ is 1 for a fermion that is a coloured triplet component and 0 otherwise.

Integration and expressing the scale dependence of $\alpha_s$ in terms of the QCD beta function results in a solution for $[I]^{-1}(t)$, whose inverse reads
\begin{equation}
  \label{eq:ift-from-t0}
  I_{f_1 f_2}(\mu) = I_{f_1 f}(\mu_0) \left[ \mathds{1}
    - \frac{1}{6 \pi^2} \ln \left( \frac{\mu}{\mu_0} \right) [I(\mu_0)]
    + \frac{1}{3 \pi^2  \beta_{s0}} \ln \left( \frac{\alpha_s(\mu)}{\alpha_s(\mu_0)} \right) [\delta_c I(\mu_0)]
  \right]^{-1}_{f f_2}\,.
\end{equation}

The above solution involves the inversion of larger matrices, that can be still performed efficiently with numerical methods.
In the following we will perform this inversion through an expansion in $I_{ff'}\ln/(6 \pi^2)$ in order to match the above result to the millicharged invariant of Eq.~(\ref{milliinvar}).
Considering fermion fields $\psi$, $\eta$ and $\chi$, defined as in section \ref{ssec:milli} we find
\begin{equation}
  \label{eq:millicharge-invariant}
  I_{\varepsilon} (\mu) \approx
  \frac{1}{6 \pi^2 I_{\psi\psi}}\sum_{h \in \eta} I_{\psi h}  I_{h \chi} \ln \left( \tfrac{m_h}{\Lmod} \right)~~.
\end{equation}

\section{ An invariant for $\tau^+\tau^-\to \g + E_{miss}$ }
\label{app:ttgg}

This Appendix discusses an   invariant  for a   process with  external gauge bosons: $\tau^+\tau^-\to A+V$.
Physically, this  would give a final state  of $\g+$ missing energy.  In the case of two massless gauge bosons,  the physical photon is the gauge boson that interacts with the SM, so this process should not occur. However,  in the case where $V^\m$
is massive,  the photon $A^\m$ is defined as the massless gauge boson, and
$\tau^+\tau^-\to \g + E_{miss}$ becomes possible.
In the case where $V^\m$
gets a mass from the vacuum expectation value  of a scalar $\phi$, this appears in the non-canonical Lagrangian (\ref{Lnc}) as
\bea
{\cal L}_{nc} \supset  \langle \phi\rangle ^2\vec{A}^\m [\vec{v}_\phi \vec{v}_\phi^T] \vec{A}^\m~~~.
\eean

The  $U(1) \times U(1)$ structure  of the two-point function  for  $V^\m$ can be written in an arbitrary basis  as
\beq
D^{\m\n}_V \propto  [K^{-1}]  \frac{[\vec{v}_\phi \vec{v}_\phi^T]}{\vec{v}_\phi\cdot [K^{-1}]  \cdot\vec{v}_\phi } [K^{-1}] ~~~,~~
[\vec{v}_\phi \vec{v}_\phi^T]=
     \left[\begin{array}{cc}
         Q_\phi^2 & Q_\phi Y_\phi\\
         Q_\phi Y_\phi&   Y^2_\phi
         \end{array}\right]
     \label{PV}
     \eeq
where the simplest field parametrisation in which to implement this matrix is to take $[K]\propto I$, and  choose the basis such that $Q_\phi=0$.
The equivalent coupling structure for the orthogonal gauge boson $A$  is orthogonal in the U(1) space, so
\beq
D^{\m\n}_A \propto  [K^{-1}]  \varepsilon \frac{[\vec{v}_\phi \vec{v}_\phi^T]}{\vec{v}_\phi\cdot \varepsilon[K^{-1}] \varepsilon \cdot\vec{v}_\phi } \varepsilon [K^{-1}] ~~~.
     \label{PA}
     \eeq
     where $\varepsilon$ is the antisymmetric 2$\times$2 matrix with unit entries.
The matrix element-squared for $|{\cal M}(\tau \tau \to A+V)|^2$ 
is then  proportional to 
 \beq
\frac{(\vec{v}_\tau \cdot [K^{-1}]\cdot \vec{v}_\phi)^2
  (\vec{v}_\tau\cdot [K^{-1}][\varepsilon]\cdot\vec{v}_\phi)^2}
     {|(\vec{v}_\phi\cdot \varepsilon[K^{-1}] \varepsilon \cdot\vec{v}_\phi)
(\vec{v}_\phi\cdot [K^{-1}]  \cdot\vec{v}_\phi)|
     } ~~.
 \label{invarttAV}
 \eeq

\end{document}